\documentclass[11pt,a4paper]{article}

\usepackage{INTERSPEECH2019}

\usepackage{graphicx}
\usepackage{url}
\usepackage{multirow}
\usepackage{cite}
\usepackage{pdfpages}

\title{Mirroring to Build Trust in Digital Assistants}

\name{Katherine Metcalf, Barry-John Theobald, Garrett Weinberg, Robert Lee, Ing-Marie Jonsson, Russ Webb, Nicholas Apostoloff}

\address{Apple Inc., 1 Apple Parkway, Cupertino, CA, 95014, USA}
\email{ \{kmetcalf, bjtheobald, gweinberg, perihare, ingmarie, rwebb, napostoloff\}@apple.com}

\date{}

\begin{document}

\maketitle

\begin{abstract}
We describe experiments towards building a conversational digital assistant that considers the preferred conversational style of the user. In particular, these experiments are designed to measure whether users prefer and trust an assistant whose conversational style matches their own. To this end we conducted a user study where subjects interacted with a digital assistant that responded in a way that either matched their conversational style, or did not. Using self-reported personality attributes and subjects' feedback on the interactions, we built models that can reliably predict a user's preferred conversational style.
\end{abstract}

\noindent\textbf{Index Terms}: speech, mirroring, conversation


\section{Introduction}\label{sec:introduction}	
Long-term reliance on digital assistants requires a sense of trust in the assistant and its abilities. Therefore, strategies for building and maintaining this trust are required, especially as digital assistants become more advanced and operate in more aspects of people's lives. 

In the context of human-human interactions, people use certain behaviors to build rapport with others \cite{delaherche2012interpersonal}. One such behavior is ``mirroring'' \cite{swaab2011early}, which occurs over the course of an interaction when people ``reflect'' some of their partner's behaviors back to them, e.g.\ adopting the posture or facial expression of the conversational partner. This phenomena, often created via the process of entrainment, has also been referred to as: mimicry, social resonance, coordination, synchrony, attunement, the chameleon effect, and so on. We hypothesize that an effective method for enhancing trust in digital assistants is for the assistant to mirror the {\it conversational style} of a user's query, specifically the degree of ``chattiness.'' We loosely define chattiness to be the degree to which a query is concise (high information density) versus talkative (low information density).

To test our hypothesis we conducted a user study, the results of which demonstrate that people not only enjoy interacting with a digital assistant that mirrors their level of chattiness in its responses, but that interacting in this fashion increases feelings of trust.  Furthermore, we demonstrate that it is possible to extract the information necessary to predict when a chatty response would be appropriate.

The rest of this paper is organized as follows: Section \ref{sec:background} discusses  background related to mirroring and building trust.  Section \ref{sec:experiment} provides an overview of the experiments, where the user study is described in Section  \ref{sec:userstudydesign} and the experiments for classifying query style are provided in Section  \ref{sec:ml}. Finally, Section \ref{sec:discussion} offers a summary and some suggested future work.


\section{Background}\label{sec:background}

People are able to engender trust and camaraderie through behavioral mirroring \cite{chartrand1999chameleon,pickering2004toward,goleman2006emotional,nenkova2008high}, where conversational partners mirror one another's interaction style as they negotiate to an agreed upon model of the world \cite{niederhoffer2002linguistic,pennebaker2003psychological,taylor2008linguistic}. Behavioral mirroring is strongly correlated with and predictive of many qualitative interaction measures \cite{chartrand1999chameleon}. It has been shown that the modalities involved in and the degree of mirroring are both predictive of how natural an interaction will be ranked \cite{nenkova2008high}. Understanding and detecting instances of mirroring has become of increasing research interest to the human computer interaction (HCI), machine learning (ML), and developmental robotics fields. Most of the work focuses on detecting and estimating the degree of non-speech-based mirroring and its effects on how people interact with one another \cite{delaherche2010multimodal,messinger2010applying,michelet2012automatic,ramseyer2011nonverbal,sun2011automatic,sun2011towards,terven2016head}. For example, the process of mirroring has been specifically leveraged to improve predictions about turn-taking in multi-person interactions. Such systems typically integrate the previous and current actions of all interactants to predict the next actions of a subset of the interactants \cite{ozkan2010latent}, e.g.\ to predict turn transitions \cite{huang2011multimodal,neiberg2011predicting} and next utterance type \cite{neiberg2011predicting}. Mirroring has also been proposed as a key tool that developmental robotics may leverage to improve language acquisition through interacting with and observing humans \cite{cangelosi2010integration}. Mirroring has since been used as a learning technique to develop social robots \cite{cangelosi2010integration,galdeano2018developmental,nakajo2018acquisition,barkan2018curiosity}. However, to the best of our knowledge, no work has explored mirroring conversational style as a behavior to be produced by a digital assistant.


\section{Experiments}\label{sec:experiment}

We describe: (1) an interactive Wizard-of-Oz (WOZ) user study, and (2) automatic prediction of preferred conversational style using the queries, responses, and participant feedback from the WOZ. In (1), all interactions between the participants and the digital assistant were strictly verbal; there was no visual realization of the digital assistant's responses.

\subsection{User Study}\label{sec:userstudydesign}

The user study evaluated the hypothesis: \emph{interacting with a digital assistant that mirrors a participant's chattiness will cause a positive change in the participant's opinion of the assistant}. In testing this, we also tested whether people who score as chatty according to our measure of interaction style ({\bf Survey 1}: Appendix~\ref{pdf:surveyone}) are more likely to prefer interacting with an assistant who is also chatty, and furthermore whether people who score as non-chatty will prefer non-chatty assistant interactions.   Prospective participants completed a questionnaire designed to assess their chattiness level along with other personality traits (e.g.\ extrovert vs.\ introvert) after volunteering, but prior to being selected for the study.   This allowed us to somewhat balance participants across \emph{types}.  After selecting the participants, they each completed a pre-study survey ({\bf Survey 2}: Appendix~\ref{pdf:surveytwo}) to determine how they use digital assistants (frequency of use, types of query, style of interaction, trustworthiness, likability, etc.).

The study consisted of three conditions: interacting with (1) a chatty, (2) a non-chatty, and (3) a mirroring digital assistant.  In all conditions the digital assistant was controlled by a wizard ({\it i.e.}\ the experimenter), and the wizard was instructed to not interact directly with the participants during the study.

In the chatty and non-chatty conditions, the participants were prompted (via instructions displayed on a wall-mounted TV display) to make verbal requests of the digital assistant for tasks in each of the following domains:  timers/alarms, calendars/reminders, navigation/directions, weather, factual information, and web search.   Prompts were text-based and deliberately kept short to limit the tendency to copy the prompts' phrasing when speaking a query.  The assistant's responses were generated for each prompt {\it a priori} and they did not vary between participants.   As an example, a prompt might read ``next meeting time'', for which the chatty response was, ``It looks like you have your next meeting after lunch at 2 P.M.''; and the corresponding non-chatty response was simply ``2 P.M.''.   After hearing a response, participants were directed to verbally classify its qualities: \emph{good}, \emph{off topic}, \emph{wrong information}, \emph{too impolite}, or \emph{too casual}, which was recorded (via a keyboard) by the wizard.  To counter presentation ordering effects, the order of the prompts was randomly assigned such that half of the participants experienced the chatty condition first, whilst the other half non-chatty. After completing both the chatty and non-chatty conditions, participants answered questions ({\bf Survey 3}: Appendix~\ref{pdf:surveythree}) about their preference for the chatty vs.\ non-chatty interactions. Participants specified their preferences both within and across the task domains.

All participants interacted with the mirroring assistant after completing both the chatty and non-chatty conditions.  The mirroring assistant interactions were designed to be as natural as possible within the confines of a WOZ user study.   The wizard for this condition was the same as was used in the chatty and non-chatty conditions, and again, the wizard was instructed to not interact with the participants during the study.  Note that in the first two conditions the wizard was not required to rate the chattiness of a query since the type of response is defined \emph{a priori} depending on the condition, and so the responses are not dependent on the conversational style of the user.  However, in this condition the role of the wizard was to assign a chattiness score (1, 2, 3, 4, or 5) for each utterance produced by a participant, which was then used to select the chattiness level of the assistant's response.

\begin{figure*}[t]
\begin{tabular}{ccc}
\includegraphics[scale=0.05]{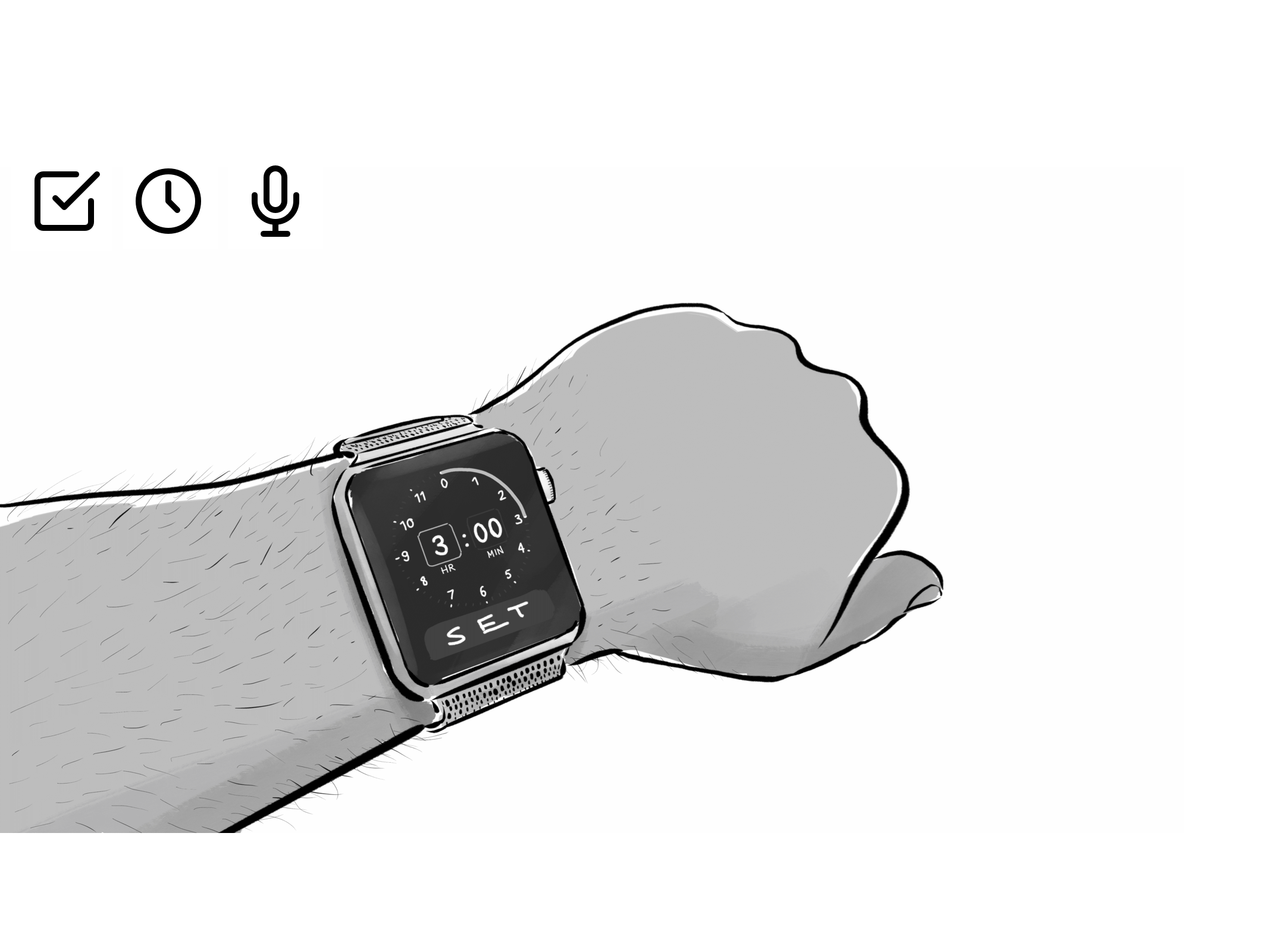} &
\includegraphics[scale=0.05]{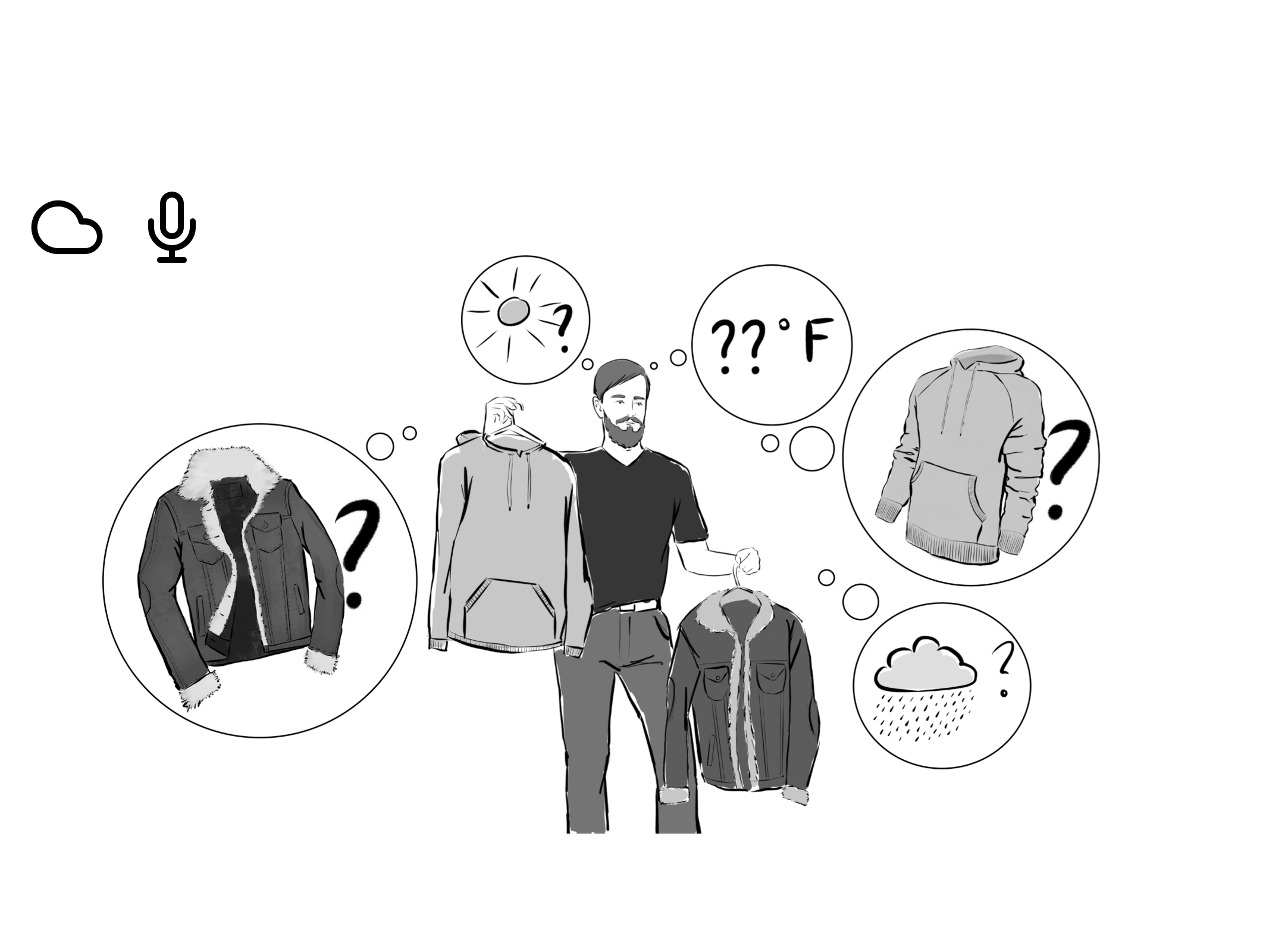} &
\includegraphics[scale=0.05]{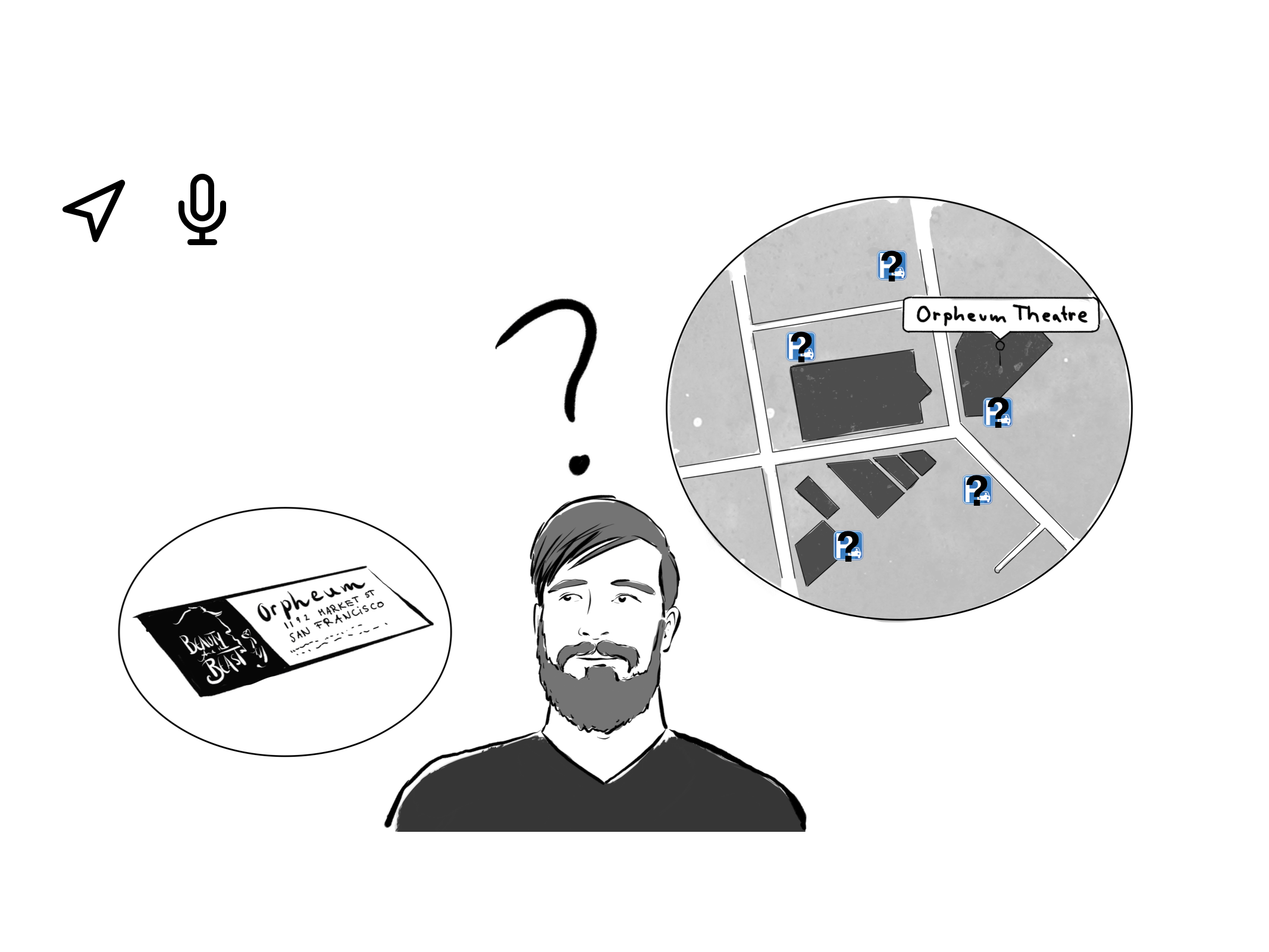} \\\vspace{5mm}
(a) set a timer for three hours & b2) find the weather at the destination & (c) get directions to the theatre \\
\includegraphics[scale=0.05]{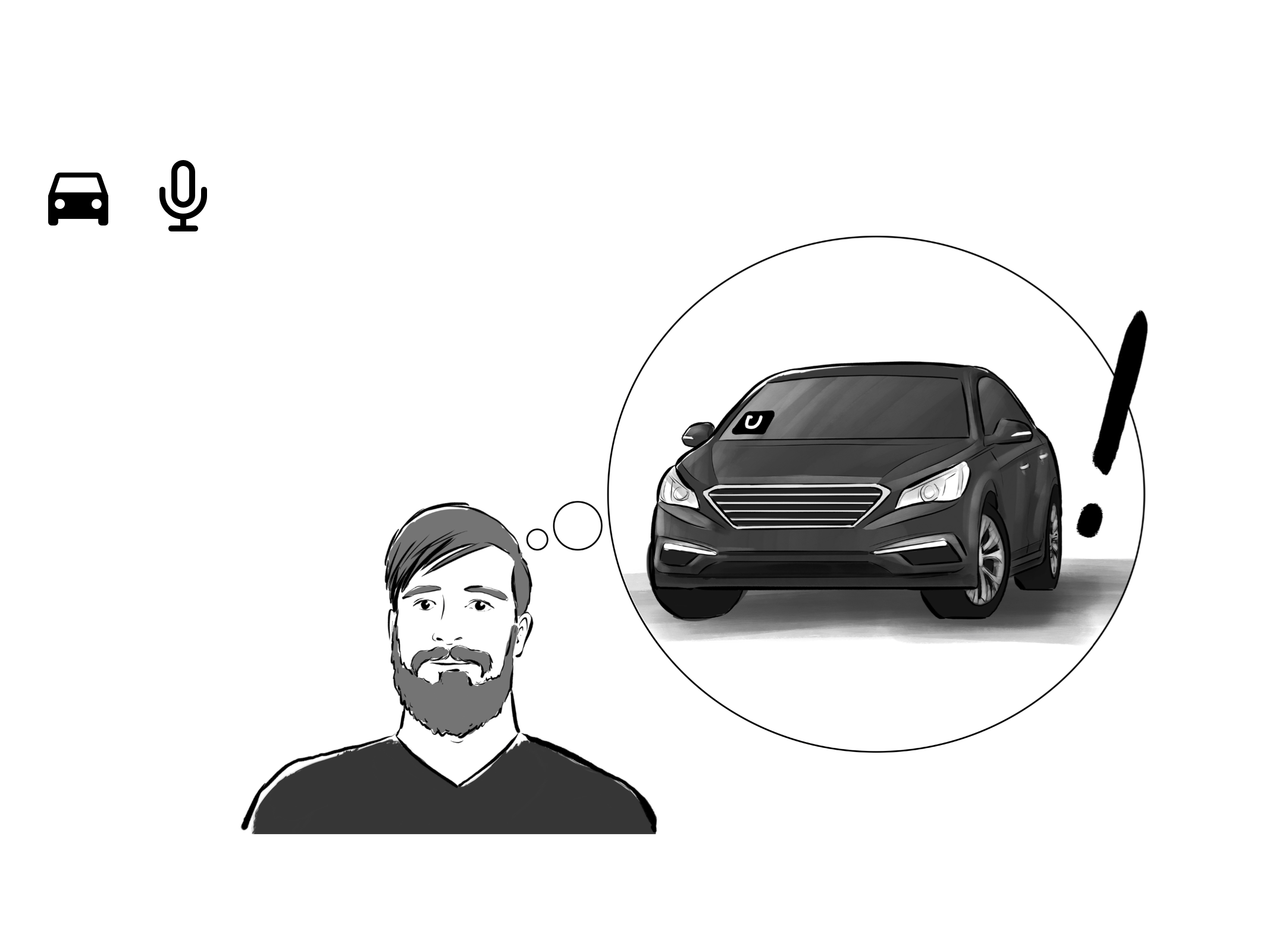} &
\includegraphics[scale=0.05]{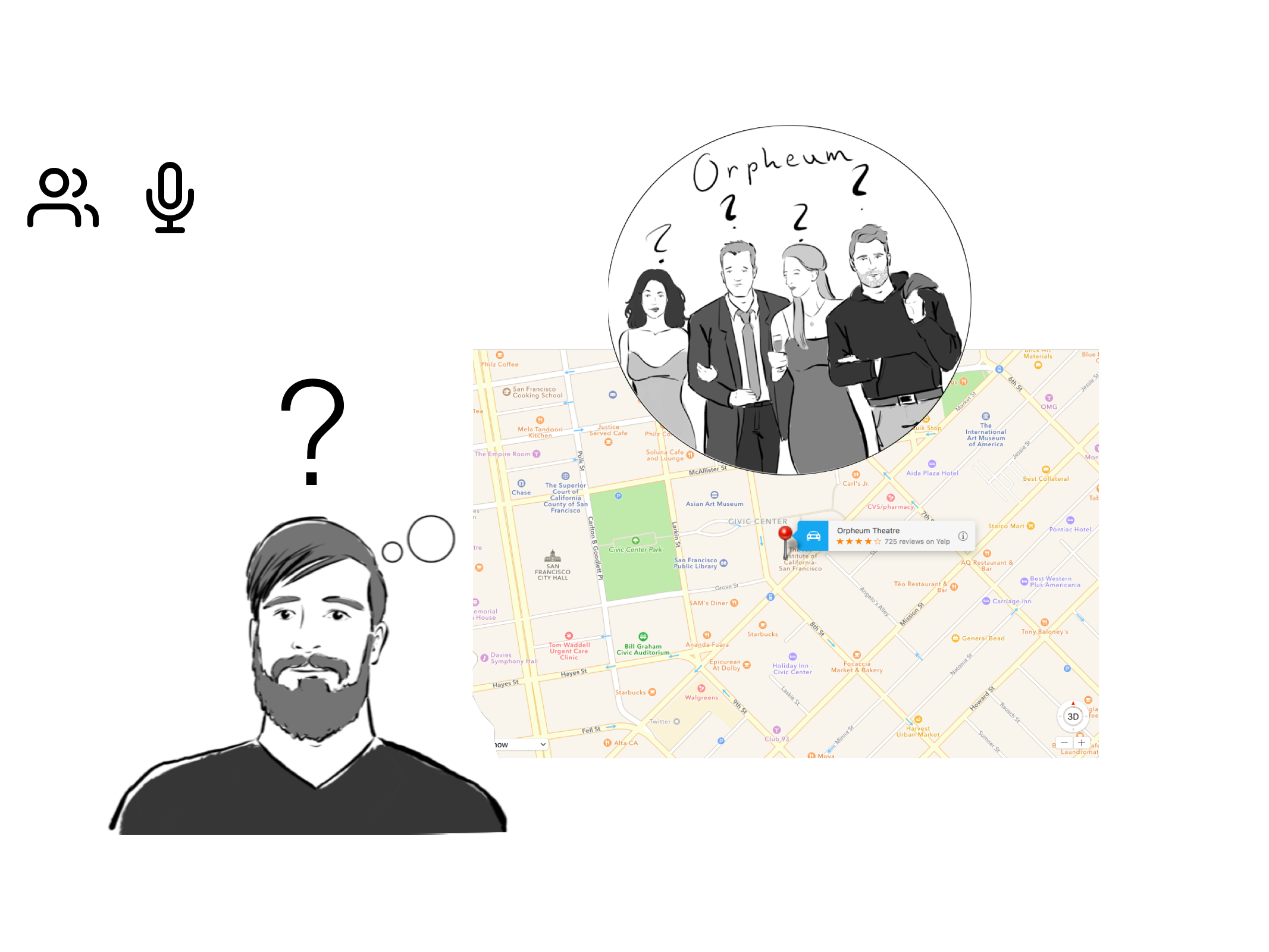} &
\includegraphics[scale=0.05]{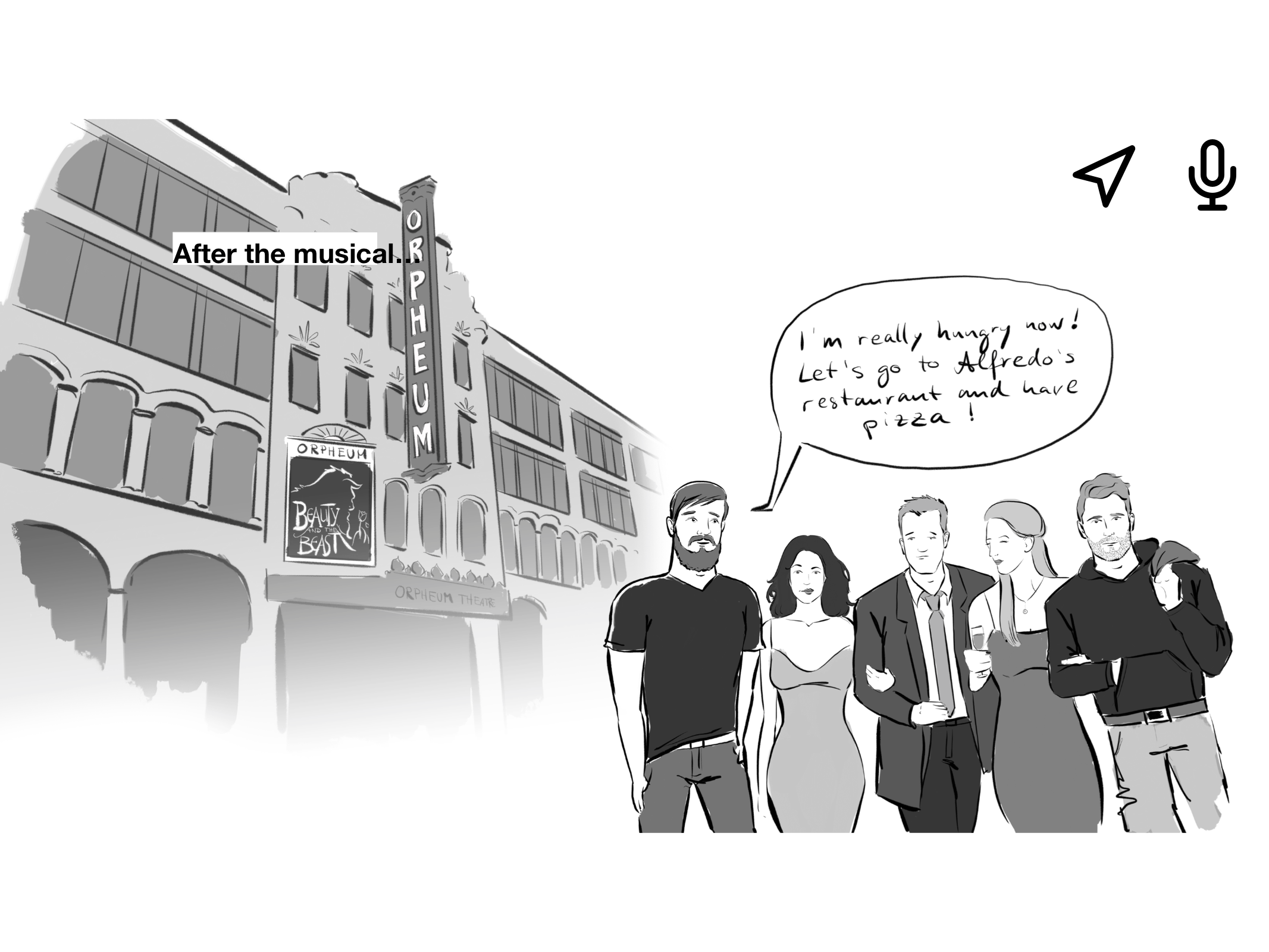} \\\vspace{5mm}
(d) call an Uber & (e) locate your friends & (f) get directions to Alfredo's \\
\multicolumn{3}{c}{\includegraphics[scale=0.05]{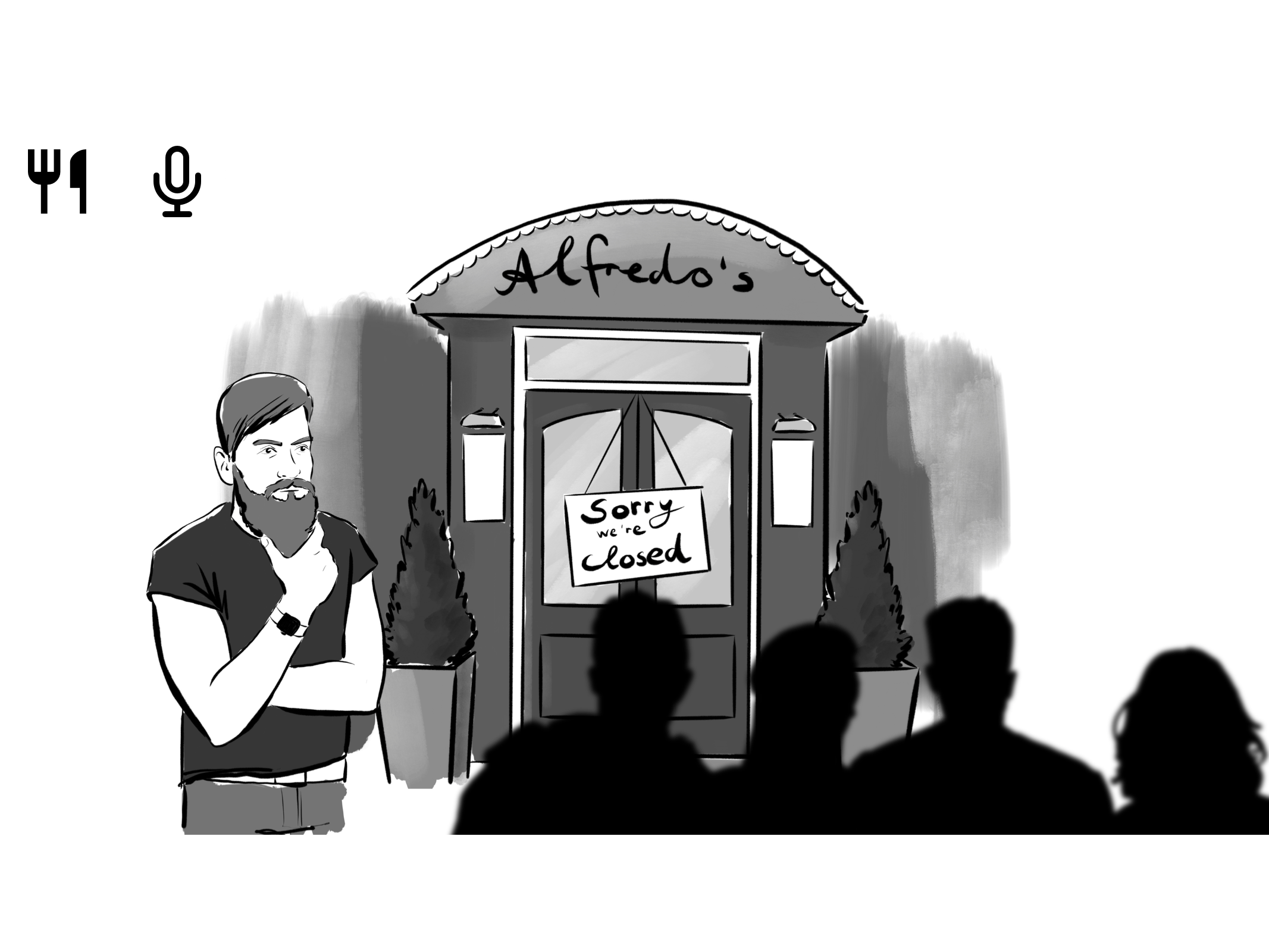}} \\
\multicolumn{3}{c}{(g) get recommendations for an alternative restaurant}
\end{tabular}
\caption{The prompts for the seven queries used in our mirroring user study.}
\label{fig:prompts}
\end{figure*}

\subsubsection{Eliciting Natural Queries for Mirrored Responses}
To guide the formation of natural inputs, participants were asked to imagine an ``evening out'' scenario, which involved meeting friends, watching a show, and planning dinner. The wizard walked subjects through the scenario, prompting them (via the TV display) to make requests using a set of image-icon pairs, see Figure \ref{fig:prompts}; no word-based prompts were used. The hand-drawn images depicted the imagined stage in the evening, and the icons indicated which of the digital assistant's task-oriented functionalities the participant should take advantage of. For example, early in the scenario walkthrough, an image of a closet of clothes was paired with a weather app icon, see Figure \ref{fig:prompts}(b). The set of possible responses associated with each prompt was fixed across participants, and the chattiness level of the selected response always matched chattiness level assigned to the query by the wizard.  The responses associated with Figure~\ref{fig:prompts}(b), sorted least to most chatty, were:
\begin{enumerate}
	\item ``74 and clear.''
	\item ``It will be 74 degrees and clear.''
	\item ``It will be a comfortable 74 degrees with sunny skies.''
	\item ``It's supposed to be 74 degrees and clear, so don't bother bringing a sweater or jacket.''
	\item ``Well, my sources are telling me that it's supposed to be 74 degrees and clear. You probably don't need to bother bringing a sweater or jacket.''
\end{enumerate}
 As in the chatty and non-chatty conditions, participants rated each of the assistant's responses. After completing all interactions in the mirroring condition, participants completed the post-study survey, which was used to measure changes in opinion about the likability and trustworthiness of the digital assistant.

\subsubsection{Setup}\label{sec:setup}

During the user study we recorded speech and video for each participant.  In particular, speech was recorded at 16 kHz mono, 16-bits per sample using a R\o de Lavalier lapel microphone and a Focusrite 2i4 USB audio interface.  Video (not used here) was recorded in both grayscale and near-infrared at resolution 1088 $\times$ 1088 at 100 frames per second, 8-bits per pixel using Ximea MQ022MG-CM and MQ022RG-CM cameras respectively.  In addition we also captured depth information (also not used here) using a PrimeSense RD1.09 sensor.

Subjects were asked to sit and face a wall-mounted TV at a distance of approximately 2.5m.  This was used to display all text and imagery to the participants, and the cameras were mounted beneath this display to obtain a good view of the participants.  The wizard sat behind a dividing screen and used a MacPro to control the digital assistant and drive the display; the same MacPro was used to synchronize all hardware and capture the data using ROS \cite{morgan2019ros}.

\begin{table}[h]
	\centering
	\begin{tabular}{ c|c|c }
		\hline
		\hline
		Task Domain & $F$-score & $p$-value \\
 		\hline
 		\hline
 		navigation/direction & 6.24 & 0.02 \\
 		\hline
 		factual information & 2.86 & 0.12 \\
 		\hline
 		timers/alarms/calendar & 0.08 & 0.79  \\
 		\hline
 		weather & 5.12 & 0.04 \\
 		\hline
 		web search & 7.73 & 0.02 \\
 		\hline
 		\hline
	\end{tabular}
	\caption{Significance of the difference in preferences for chatty vs.\ non-chatty responses per domain. There is no significant difference for timers and factual information.}
	\label{tab:domainstatsig}
\end{table}

\begin{table*}[t]
	\centering
	\begin{tabular}{ c|c|c|c|c|c|c|c|c|c|c }
		\hline
		\hline
		\multicolumn{11}{c}{\textbf{Leave One Participant Out}} \\
		\hline
		\hline
		& \textit{Log.} & \textit{Naive} & \multirow{2}{*}{\textit{ANN}} & \textit{RF} & \textit{RF} & \multirow{2}{*}{\textit{SVM}} & \textit{SVM} & \textit{SVM} & \textit{SVM} & \textit{SVM} \\
		& \textit{Reg.} & \textit{Bayes} & & \textit{(Gini)} & \textit{(Entropy)} & & \textit{(RBF)} & \textit{(Linear)} & \textit{(Poly.)} & \textit{(Sig.)} \\
 		\hline
 		\hline
 		Detector & 0.60 & 0.71 & 0.82 & 0.83 & 0.83 & 0.83 & 0.82 & 0.84 & 0.82 & 0.84 \\
 		\hline
 		Selector & 0.65 & 0.71 & 0.79 & 0.79 & 0.81 & 0.83 & 0.84 & 0.81 & 0.83 & 0.84 \\
 		\hline
 		\hline
 		\multicolumn{11}{c}{\textbf{80/20 Split}} \\
		\hline
		\hline
		 Detector & 0.62 & 0.74 & 0.85 & 0.85 & 0.85 & 0.86 & 0.86 & 0.86 & 0.85 & 0.86 \\
 		\hline
 		Selector & 0.68 & 0.78 & 0.82 & 0.81 & 0.85 & 0.85 & 0.87 & 0.84 & 0.86 & 0.86 \\
 		\hline
 		\hline
	\end{tabular}
	\caption{$F_{1}$-scores for the utterance chattiness and response chattiness preference binary classification tasks (chatty vs.\ non-chatty) for both the Leave One Participant Out (top) and the 80/20 evaluation splits (bottom).}
	\label{tab:mlresults}
\end{table*}

\subsubsection{Results and Discussion}\label{sec:userstudyresultsdiscussion}

In total twenty people (three women and seventeen men) participated in the study, with session durations ranging from 17 to 56 minutes, depending on participants' verbosity. The majority of participants (70\%) preferred interacting with the chatty digital assistant. According to participant responses to the personality and interaction style survey (\textbf{Survey 1}), 60\% of participants were generally chatty and 40\% were non-chatty. As more people preferred the chatty response than were identified as chatty, one's own self-reported conversational style does not necessarily predict one's preference for assistant chattiness. However, in general, the participants identified as chatty preferred the chatty interactions, and those identified as non-chatty preferred the non-chatty interactions.

The effect of mirroring on opinions of likability and trustworthiness was tested using a one-way ANOVA. We compared the participants' trustworthiness ratings of the assistant from the pre-study (mean=4.0, stdev=0.48) and post-study (mean=4.46, stdev=0.31) surveys.  Users were asked to rate how much they agreed with the statements about liking vs.\ not liking interacting with the assistant on a seven-point Likert scale, where one corresponds to strong disagreement, four is neutral, and seven corresponds to strong agreement (see {\bf Survey 1}). The difference in the mean score pre-mirroring and post-mirroring conditions was statistically significant ($\textnormal{f-score}=7.12$, ${\it p} \leq 0.01$), meaning that interacting with a style-mirroring assistant had a significant, positive impact on opinions and trustworthiness. Additionally, the task domain had a smaller, but still significant impact on whether subjects preferred chatty or non-chatty responses ($\textnormal{f-score}=2.67$, ${\it p} \leq 0.02$). For a specific breakdown by task domain, see Table~\ref{tab:domainstatsig}.

Anecdotal evidence from comments in the post-study debrief suggest that participants prefer the assistant in the mirroring conditions.  In summary, we conclude that chattiness preferences differ across individuals and across task domains, but mirroring user chattiness increases feelings of likability and trustworthiness in digital assistants.  Given the positive impact of mirroring chattiness on interaction, we proceeded to build classifiers to determine whether features extracted from user speech could be used to estimate their level of chattiness, and thus the appropriate chattiness level of a response.


\section{Detecting Preferred Interaction Style}\label{sec:ml}


We built multi-speaker and speaker-independent classifiers to detect from a query utterance:  (1) if the utterance is chatty or non-chatty, and (2) if a chatty vs.\ non-chatty response would be preferred.  The chatty or not classification was based solely on the audio features and did not include a measure of utterance duration as duration does not have a strong enough correlation with information density (our measure of utterance chattiness). The chatty vs.\ non-chatty target label for each utterance was extracted from the survey responses, overall participant chatty vs.\ non-chatty labels were extracted from {\bf Survey 1}, and response preference labels (\emph{good}, \emph{too casual}, etc.) were extracted from the digital assistant response evaluations obtained in the user-study. Only the utterances from the chatty and non-chatty conditions were included in the data set. Each utterance was assigned two labels, where one indicated whether the speaker was chatty, and the other indicated the response preference for the utterance. From the speech, 95 acoustic features were extracted: the mean, standard deviation, maximum, minimum of the fundamental frequency, energy, the first twelve MFCC's, and the first five formants \cite{jurafsky2009extracting}.

Ten classifiers were trained on the two binary classification tasks: logistic regression, naive Bayes, artificial neural network (one hidden layer with 64 units), random forest using Gini, random forest using entropy, SVM, SVM with an RBF, polynomial, linear, and sigmoid kernel. These classifiers were selected for their simplicity due to the small number of data points, and we used the standard implementations in \texttt{scikit-learn} \cite{sklearn}. The classifiers were evaluated with both an 80/20 split of training/test data, so we train on samples for all speakers and test on different data from the same speakers (multi-speaker), and a leave-one-participant-out train/test split (speaker-independent). Performance was evaluated according to the $F_{1}$-score due to the label imbalance.

\subsubsection{Results and Discussion}\label{mlexperiments}

The classification results are shown in Table \ref{tab:mlresults}, which demonstrate that the classifiers performed well for both forms of evaluation split. This is a promising indicator that both a speaker's degree of chattiness and their preference for chatty vs.\ non-chatty response can be detected reliably. The majority of classifiers had performance comparable to one another and better than chance, with the SVM methods performing best on both types of evaluation split and both classification tasks. These results are encouraging, especially given the small size of the data set. Performance on the 80/20 split indicates that the algorithms do not need many samples from a user to learn. Performance on the leave-one-participant-out split suggests that the models are able to generalize to new speakers. However, access to samples across participants does improve performance for all the classifiers tested.


\section{Conclusion and Future Directions}\label{sec:discussion}

We have shown that user opinion of the likability and trustworthiness of a digital assistant improves when the assistant mirrors the degree of chattiness of the user, and that the information necessary to accomplish this mirroring can be extracted from user speech. Future work will investigate detecting ranges of chattiness rather than the binary labels used here, expand the participant pool, and we will use multimodal signals from the videos and depth images to measure the degree to which users appreciate the assistant responses.

\bibliography{bibliography}

\begin{thebibliography}{10}
\providecommand{\url}[1]{#1}
\csname url@samestyle\endcsname
\providecommand{\newblock}{\relax}
\providecommand{\bibinfo}[2]{#2}
\providecommand{\BIBentrySTDinterwordspacing}{\spaceskip=0pt\relax}
\providecommand{\BIBentryALTinterwordstretchfactor}{4}
\providecommand{\BIBentryALTinterwordspacing}{\spaceskip=\fontdimen2\font plus
\BIBentryALTinterwordstretchfactor\fontdimen3\font minus
  \fontdimen4\font\relax}
\providecommand{\BIBforeignlanguage}[2]{{%
\expandafter\ifx\csname l@#1\endcsname\relax
\typeout{** WARNING: IEEEtran.bst: No hyphenation pattern has been}%
\typeout{** loaded for the language `#1'. Using the pattern for}%
\typeout{** the default language instead.}%
\else
\language=\csname l@#1\endcsname
\fi
#2}}
\providecommand{\BIBdecl}{\relax}
\BIBdecl

\bibitem{delaherche2012interpersonal}
E.~Delaherche, M.~Chetouani, A.~Mahdhaoui, C.~Saint-Georges, S.~Viaux, and
  D.~Cohen, ``Interpersonal synchrony: A survey of evaluation methods across
  disciplines,'' \emph{IEEE Transactions on Affective Computing}, vol.~3,
  no.~3, pp. 349--365, 2012.

\bibitem{swaab2011early}
R.~I. Swaab, W.~W. Maddux, and M.~Sinaceur, ``Early words that work: When and
  how virtual linguistic mimicry facilitates negotiation outcomes,''
  \emph{Journal of Experimental Social Psychology}, vol.~47, no.~3, pp.
  616--621, 2011.

\bibitem{chartrand1999chameleon}
T.~Chartrand and J.~Bargh, ``The chameleon effect: the perception--behavior
  link and social interaction.'' \emph{Journal of personality and social
  psychology}, vol.~76, no.~6, p. 893, 1999.

\bibitem{pickering2004toward}
M.~Pickering and S.~Garrod, ``Toward a mechanistic psychology of dialogue,''
  \emph{Behavioral and brain sciences}, vol.~27, no.~2, pp. 169--190, 2004.

\bibitem{goleman2006emotional}
D.~Goleman, \emph{Emotional intelligence}.\hskip 1em plus 0.5em minus
  0.4em\relax Bantam, 2006.

\bibitem{nenkova2008high}
A.~Nenkova, A.~Gravano, and J.~Hirschberg, ``High frequency word entrainment in
  spoken dialogue,'' in \emph{Proceedings of the 46th annual meeting of the
  association for computational linguistics on human language technologies:
  Short papers}.\hskip 1em plus 0.5em minus 0.4em\relax Association for
  Computational Linguistics, 2008, pp. 169--172.

\bibitem{niederhoffer2002linguistic}
K.~Niederhoffer and J.~Pennebaker, ``Linguistic style matching in social
  interaction,'' \emph{Journal of Language and Social Psychology}, vol.~21,
  no.~4, pp. 337--360, 2002.

\bibitem{pennebaker2003psychological}
J.~Pennebaker, M.~Mehl, and K.~Niederhoffer, ``Psychological aspects of natural
  language use: Our words, our selves,'' \emph{Annual review of psychology},
  vol.~54, no.~1, pp. 547--577, 2003.

\bibitem{taylor2008linguistic}
P.~Taylor and S.~Thomas, ``Linguistic style matching and negotiation outcome,''
  \emph{Negotiation and Conflict Management Research}, vol.~1, no.~3, pp.
  263--281, 2008.

\bibitem{delaherche2010multimodal}
E.~Delaherche and M.~Chetouani, ``Multimodal coordination: exploring relevant
  features and measures,'' in \emph{Proceedings of the 2nd international
  workshop on Social signal processing}.\hskip 1em plus 0.5em minus 0.4em\relax
  ACM, 2010, pp. 47--52.

\bibitem{messinger2010applying}
D.~Messinger, P.~Ruvolo, N.~Ekas, and A.~Fogel, ``Applying machine learning to
  infant interaction: The development is in the details,'' \emph{Neural
  Networks}, vol.~23, no. 8-9, pp. 1004--1016, 2010.

\bibitem{michelet2012automatic}
S.~Michelet, K.~Karp, E.~Delaherche, C.~Achard, and M.~Chetouani, ``Automatic
  imitation assessment in interaction,'' in \emph{International Workshop on
  Human Behavior Understanding}.\hskip 1em plus 0.5em minus 0.4em\relax
  Springer, 2012, pp. 161--173.

\bibitem{ramseyer2011nonverbal}
F.~Ramseyer and W.~Tschacher, ``Nonverbal synchrony in psychotherapy:
  coordinated body movement reflects relationship quality and outcome.''
  \emph{Journal of consulting and clinical psychology}, vol.~79, no.~3, p. 284,
  2011.

\bibitem{sun2011automatic}
X.~Sun, A.~Nijholt, K.~Truong, and M.~Pantic, ``Automatic visual mimicry
  expression analysis in interpersonal interaction,'' in \emph{CVPR 2011
  WORKSHOPS}.\hskip 1em plus 0.5em minus 0.4em\relax IEEE, 2011, pp. 40--46.

\bibitem{sun2011towards}
X.~Sun, K.~Truong, M.~Pantic, and A.~Nijholt, ``Towards visual and vocal
  mimicry recognition in human-human interactions,'' in \emph{2011 IEEE
  International Conference on Systems, Man, and Cybernetics}.\hskip 1em plus
  0.5em minus 0.4em\relax IEEE, 2011, pp. 367--373.

\bibitem{terven2016head}
J.~Terven, B.~Raducanu, M.~Meza-de Luna, and J.~Salas, ``Head-gestures
  mirroring detection in dyadic social interactions with computer vision-based
  wearable devices,'' \emph{Neurocomputing}, vol. 175, pp. 866--876, 2016.

\bibitem{ozkan2010latent}
D.~Ozkan, K.~Sagae, and L.~Morency, ``Latent mixture of discriminative experts
  for multimodal prediction modeling,'' in \emph{Proceedings of the 23rd
  International Conference on Computational Linguistics}.\hskip 1em plus 0.5em
  minus 0.4em\relax Association for Computational Linguistics, 2010, pp.
  860--868.

\bibitem{huang2011multimodal}
L.~Huang, L.~Morency, and J.~Gratch, ``A multimodal end-of-turn prediction
  model: learning from parasocial consensus sampling,'' in \emph{The 10th
  International Conference on Autonomous Agents and Multiagent Systems-Volume
  3}.\hskip 1em plus 0.5em minus 0.4em\relax International Foundation for
  Autonomous Agents and Multiagent Systems, 2011, pp. 1289--1290.

\bibitem{neiberg2011predicting}
D.~Neiberg and J.~Gustafson, ``Predicting speaker changes and listener
  responses with and without eye-contact,'' in \emph{Twelfth Annual Conference
  of the International Speech Communication Association}, 2011.

\bibitem{cangelosi2010integration}
A.~Cangelosi, G.~Metta, G.~Sagerer, S.~Nolfi, C.~Nehaniv, K.~Fischer, J.~Tani,
  T.~Belpaeme, G.~Sandini, F.~Nori, and L.~Fadiga, ``Integration of action and
  language knowledge: A roadmap for developmental robotics,'' \emph{IEEE
  Transactions on Autonomous Mental Development}, vol.~2, no.~3, pp. 167--195,
  2010.

\bibitem{galdeano2018developmental}
A.~Galdeano, A.~Gonnot, C.~Cottet, S.~Hassas, M.~Lefort, and A.~Cordier,
  ``Developmental learning for social robots in real-world interactions,'' in
  \emph{First Workshop on Social Robots in the Wild at the 13th Annual ACM/IEEE
  International Conference on Human-Robot Interaction (HRI 2018)}, 2018.

\bibitem{nakajo2018acquisition}
R.~Nakajo, S.~Murata, H.~Arie, and T.~Ogata, ``Acquisition of viewpoint
  transformation and action mappings via sequence to sequence imitative
  learning by deep neural networks,'' \emph{Frontiers in neurorobotics},
  vol.~12, 2018.

\bibitem{barkan2018curiosity}
\BIBentryALTinterwordspacing
J.~B. and G.~G., ``Deep curiosity loops in social environments,'' \emph{CoRR},
  vol. abs/1806.03645, 2018. [Online]. Available:
  \url{http://arxiv.org/abs/1806.03645}
\BIBentrySTDinterwordspacing

\bibitem{morgan2019ros}
M.~Q., K.~Conley, B.~Gerkey, J.~Faust, T.~Foote, J.~Leibs, R.~Wheeler, and
  A.~Ng, ``{ROS}: an open-source robot operating system,'' in \emph{ICRA
  Workshop on Open Source Software}, 2009.

\bibitem{jurafsky2009extracting}
D.~Jurafsky, R.~Ranganath, and D.~McFarland, ``Extracting social meaning:
  Identifying interactional style in spoken conversation,'' in
  \emph{Proceedings of Human Language Technologies: The 2009 Annual Conference
  of the North American Chapter of the Association for Computational
  Linguistics}.\hskip 1em plus 0.5em minus 0.4em\relax Association for
  Computational Linguistics, 2009, pp. 638--646.

\bibitem{sklearn}
\BIBentryALTinterwordspacing
F.~Pedregosa, G.~Varoquaux, A.~Gramfort, V.~Michel, B.~Thirion, O.~Grisel,
  M.~Blondel, P.~Prettenhofer, R.~Weiss, V.~Dubourg, J.~Vanderplas, A.~Passos,
  D.~Cournapeau, M.~Brucher, M.~Perrot, and E.~Duchesnay, ``Scikit-learn:
  Machine learning in python,'' \emph{J. Mach. Learn. Res.}, vol.~12, pp.
  2825--2830, Nov. 2011. [Online]. Available:
  \url{http://dl.acm.org/citation.cfm?id=1953048.2078195}
\BIBentrySTDinterwordspacing

\end{thebibliography}
\bibliographystyle{IEEEtran}

\section{Appendix}

\newpage
\subsection{Survey 1} \label{pdf:surveyone}
\includepdf[scale=.75]{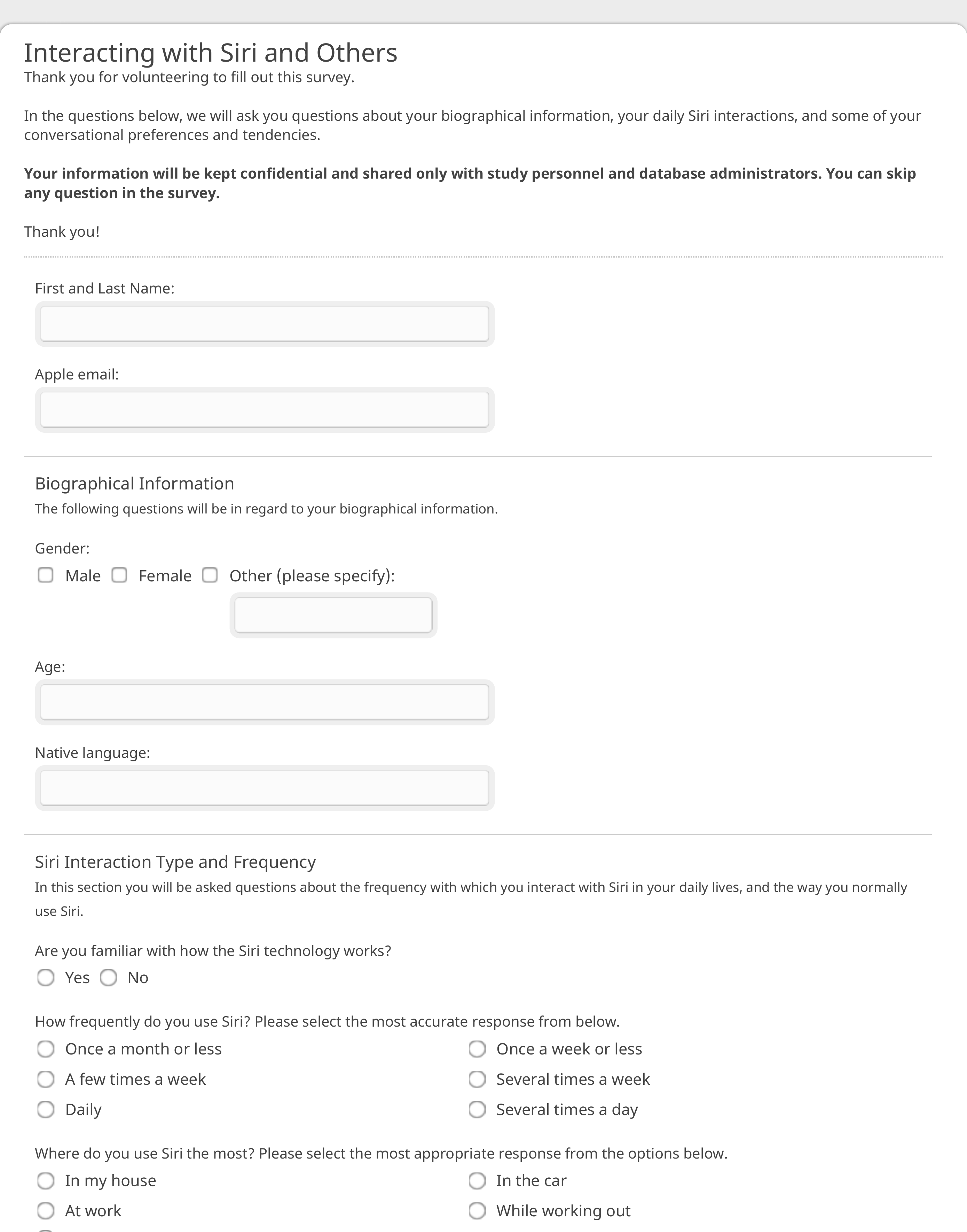}

\newpage
\subsection{Survey 2} \label{pdf:surveytwo}
\includepdf[scale=.75]{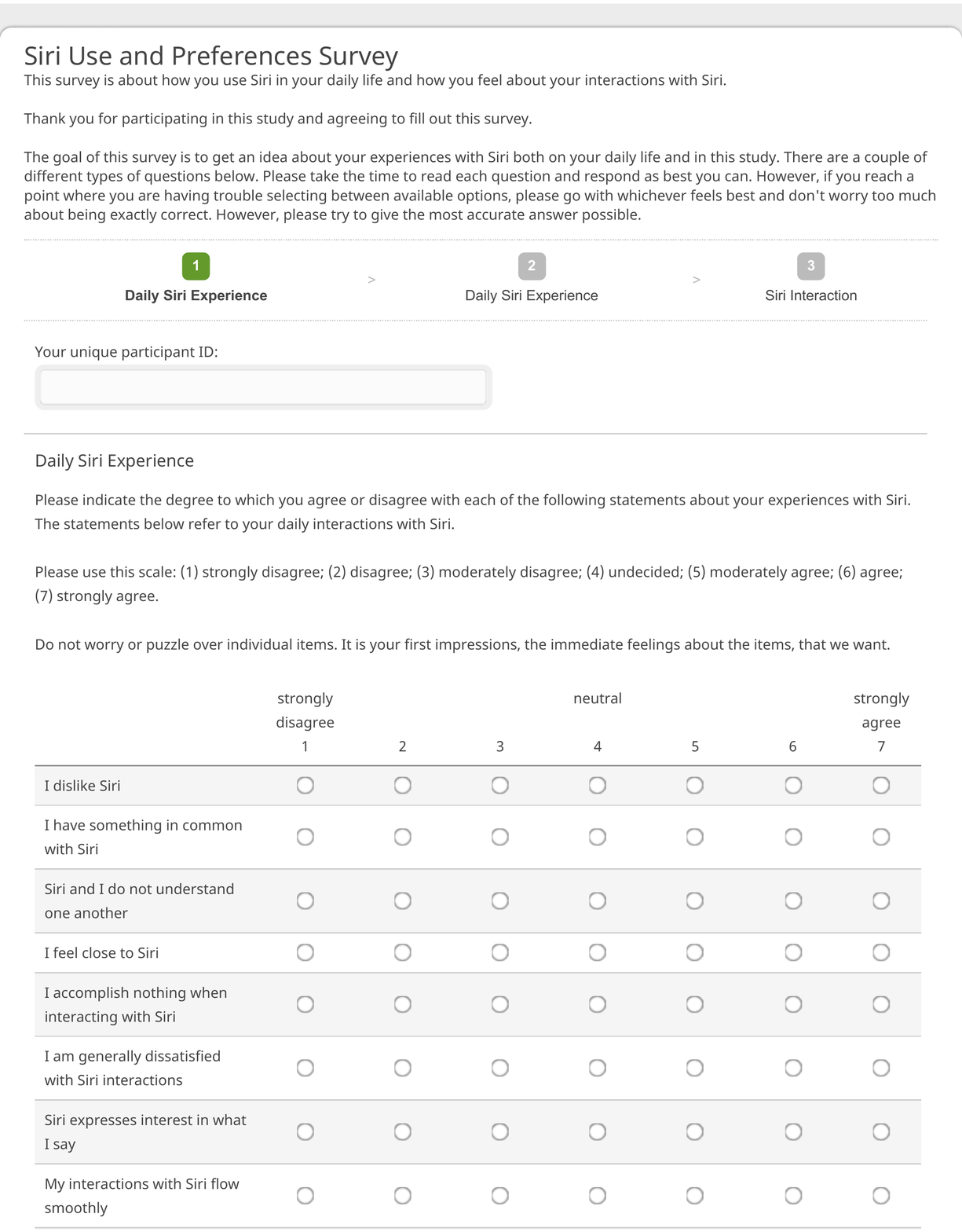}
\newpage
\includepdf[scale=.75]{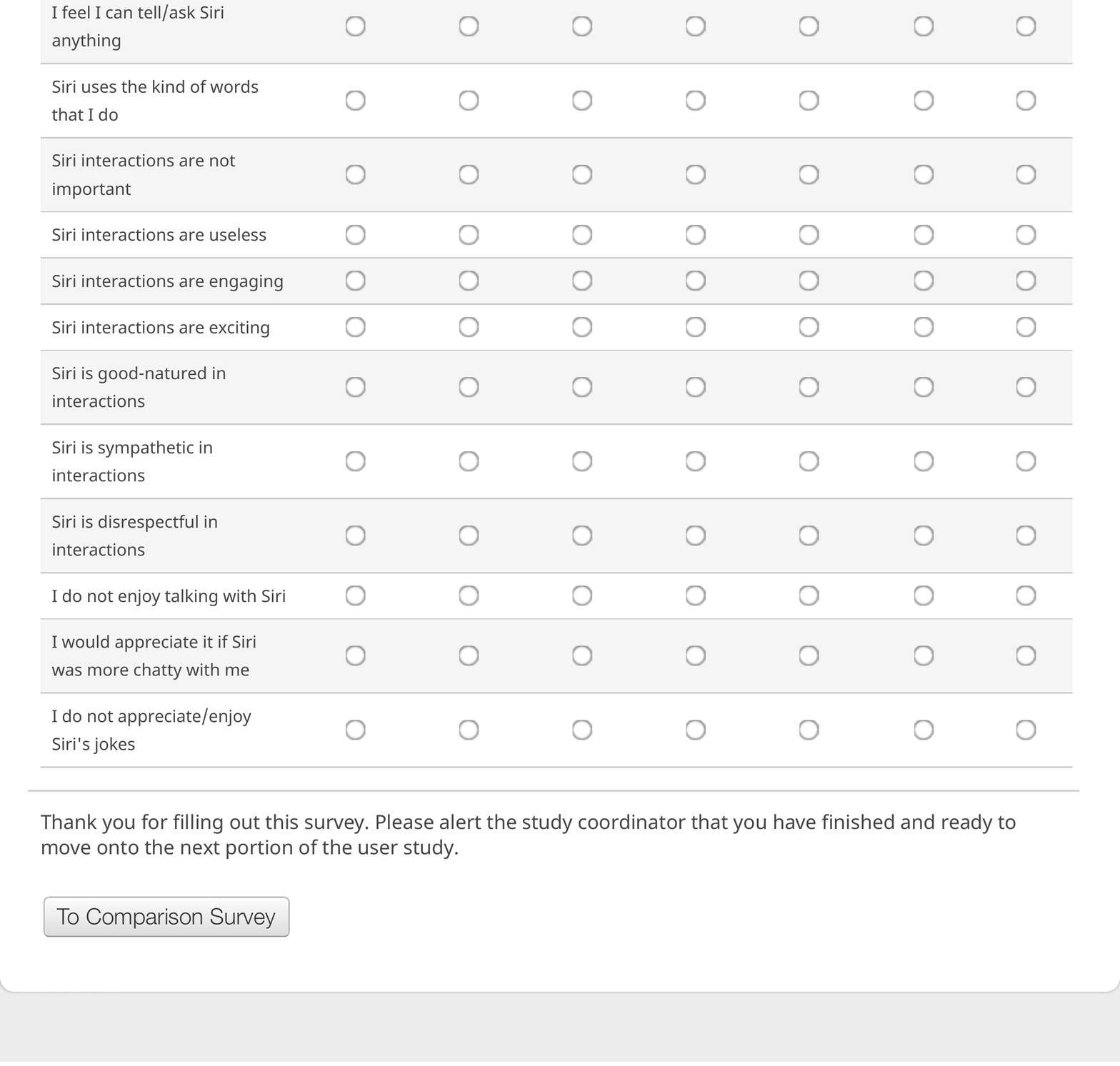}

\newpage
\subsection{Survey 3} \label{pdf:surveythree}


\section*{Supplementary material}

## 7.1. Survey 1

### Interacting with Siri and Others

Thank you for volunteering to fill out this survey.

In the questions below, we will ask you questions about your biographical information, your daily Siri interactions, and some of your conversational preferences and tendencies.

**Your information will be kept confidential and shared only with study personnel and database administrators. You can skip any question in the survey.**

Thank you!

First and Last Name:

Apple email:

### Biographical Information

The following questions will be in regard to your biographical information.

Gender:

☐ Male  ☐ Female  ☐ Other (please specify):

Age:

Native language:

### Siri Interaction Type and Frequency

In this section you will be asked questions about the frequency with which you interact with Siri in your daily lives, and the way you normally use Siri.

Are you familiar with how the Siri technology works?

○ Yes  ○ No

How frequently do you use Siri? Please select the most accurate response from below.

○ Once a month or less  ○ Once a week or less
○ A few times a week    ○ Several times a week
○ Daily                 ○ Several times a day

Where do you use Siri the most? Please select the most appropriate response from the options below.

○ In my house  ○ In the car
○ At work      ○ While working out

## 7.2. Survey 2

# Siri Use and Preferences Survey

This survey is about how you use Siri in your daily life and how you feel about your interactions with Siri.

Thank you for participating in this study and agreeing to fill out this survey.

The goal of this survey is to get an idea about your experiences with Siri both on your daily life and in this study. There are a couple of different types of questions below. Please take the time to read each question and respond as best you can. However, if you reach a point where you are having trouble selecting between available options, please go with whichever feels best and don't worry too much about being exactly correct. However, please try to give the most accurate answer possible.

| **1** | > | 2 | > | 3 |
|---|---|---|---|---|
| **Daily Siri Experience** | | Daily Siri Experience | | Siri Interaction |

Your unique participant ID:

### Daily Siri Experience

Please indicate the degree to which you agree or disagree with each of the following statements about your experiences with Siri. The statements below refer to your daily interactions with Siri.

Please use this scale: (1) strongly disagree; (2) disagree; (3) moderately disagree; (4) undecided; (5) moderately agree; (6) agree; (7) strongly agree.

Do not worry or puzzle over individual items. It is your first impressions, the immediate feelings about the items, that we want.

|  | strongly disagree 1 | 2 | 3 | neutral 4 | 5 | 6 | strongly agree 7 |
|---|---|---|---|---|---|---|---|
| I dislike Siri | ○ | ○ | ○ | ○ | ○ | ○ | ○ |
| I have something in common with Siri | ○ | ○ | ○ | ○ | ○ | ○ | ○ |
| Siri and I do not understand one another | ○ | ○ | ○ | ○ | ○ | ○ | ○ |
| I feel close to Siri | ○ | ○ | ○ | ○ | ○ | ○ | ○ |
| I accomplish nothing when interacting with Siri | ○ | ○ | ○ | ○ | ○ | ○ | ○ |
| I am generally dissatisfied with Siri interactions | ○ | ○ | ○ | ○ | ○ | ○ | ○ |
| Siri expresses interest in what I say | ○ | ○ | ○ | ○ | ○ | ○ | ○ |
| My interactions with Siri flow smoothly | ○ | ○ | ○ | ○ | ○ | ○ | ○ |

| | | | | | | | |
|---|---|---|---|---|---|---|---|
| I feel I can tell/ask Siri anything | ○ | ○ | ○ | ○ | ○ | ○ | ○ |
| Siri uses the kind of words that I do | ○ | ○ | ○ | ○ | ○ | ○ | ○ |
| Siri interactions are not important | ○ | ○ | ○ | ○ | ○ | ○ | ○ |
| Siri interactions are useless | ○ | ○ | ○ | ○ | ○ | ○ | ○ |
| Siri interactions are engaging | ○ | ○ | ○ | ○ | ○ | ○ | ○ |
| Siri interactions are exciting | ○ | ○ | ○ | ○ | ○ | ○ | ○ |
| Siri is good-natured in interactions | ○ | ○ | ○ | ○ | ○ | ○ | ○ |
| Siri is sympathetic in interactions | ○ | ○ | ○ | ○ | ○ | ○ | ○ |
| Siri is disrespectful in interactions | ○ | ○ | ○ | ○ | ○ | ○ | ○ |
| I do not enjoy talking with Siri | ○ | ○ | ○ | ○ | ○ | ○ | ○ |
| I would appreciate it if Siri was more chatty with me | ○ | ○ | ○ | ○ | ○ | ○ | ○ |
| I do not appreciate/enjoy Siri's jokes | ○ | ○ | ○ | ○ | ○ | ○ | ○ |

Thank you for filling out this survey. Please alert the study coordinator that you have finished and ready to move onto the next portion of the user study.

To Comparison Survey

## 7.3. Survey 3

### Siri Use and Preferences Survey

This survey is about how you use Siri in your daily life and how you feel about your interactions with Siri.

Thank you for participating in this study and agreeing to fill out this survey.

The goal of this survey is to get an idea about your experiences with Siri both on your daily life and in this study. There are a couple of different types of questions below. Please take the time to read each question and respond as best you can. However, if you reach a point where you are having trouble selecting between available options, please go with whichever feels best and don't worry too much about being exactly correct. However, please try to give the most accurate answer possible.

| **1** | > | **2** | > | **3** |
|---|---|---|---|---|
| Daily Siri Experience | | **Daily Siri Experience** | | Siri Interaction |

#### Siri Interactions Comparison

The goal for the questions below is to allow us to understand how you feel about the two Siri interactions you just had and to get an idea about your preferences between the two.

Each of the questions below asks you about the interaction during which you enjoyed completing the specified actions with Siri.

Overall, my preferred interaction with Siri for all types of queries was the _____.
◯ first interaction  ◯ second interaction  ◯ neither (I liked both equally)

My preferred interaction with Siri for setting timers, alarms, and reminders was the _____.
◯ first interaction  ◯ second interaction  ◯ neither (I liked both equally)

My preferred interaction with Siri for searching for or navigating to places was the _____.
◯ first interaction  ◯ second interaction  ◯ neither (I liked both equally)

My preferred interaction with Siri for asking about the weather was the _____.
◯ first interaction  ◯ second interaction  ◯ neither (I liked both equally)

My preferred interaction with Siri for just chatting or seeing how Siri will respond to something was the _____.
◯ first interaction  ◯ second interaction  ◯ neither (I liked both equally)

My preferred interaction with Siri for asking about factual or historic information was the _____.
◯ first interaction  ◯ second interaction  ◯ neither (I liked both equally)

My preferred interaction with Siri for search the web was the _____.
◯ first interaction  ◯ second interaction  ◯ neither (I liked both equally)

[ To Post Condition Survey ]  Previous



\end{document}